\documentclass[aip,amsmath,amssymb,reprint]{revtex4-1}

\usepackage{graphicx}
\usepackage{dcolumn}
\usepackage{bm}
\usepackage{textcomp}

\usepackage[utf8]{inputenc}
\usepackage[T1]{fontenc}
\usepackage[font={normal, bf}]{subcaption}

\usepackage{overpic}
\graphicspath{ {./} }
\usepackage{ulem}
\usepackage{xcolor}

\draft 

\begin{document}

\title{Monolithic Integration of Superconducting-Nanowire Single-Photon Detectors with Josephson Junctions for Scalable Single-photon Sensing}

\author{Saeed Khan}
\email[]{saeed.khan@nist.gov}
\affiliation{National Institute of Standards and Technology, 325 Broadway, Boulder, Colorado 80305, USA \looseness = -1}
\affiliation{Department of Physics, University of Colorado Boulder, 390 UCB, Boulder, CO, USA, 80309\looseness = -1}

\author{Bryce A. Primavera}
\affiliation{National Institute of Standards and Technology, 325 Broadway, Boulder, Colorado 80305, USA \looseness = -1}
\affiliation{Department of Physics, University of Colorado Boulder, 390 UCB, Boulder, CO, USA, 80309\looseness = -1}

\author{Richard P. Mirin}
\affiliation{National Institute of Standards and Technology, 325 Broadway, Boulder, Colorado 80305, USA \looseness = -1}

\author{Sae Woo Nam}
\affiliation{National Institute of Standards and Technology, 325 Broadway, Boulder, Colorado 80305, USA \looseness = -1}

\author{Jeffrey M. Shainline}
\affiliation{National Institute of Standards and Technology, 325 Broadway, Boulder, Colorado 80305, USA \looseness = -1}

\date{\today}
\begin{abstract}
We demonstrate superconducting single-photon detectors that integrate signals locally at each pixel. This capability is realized by the monolithic integration of superconducting-nanowire single-photon detectors with Josephson electronics. The motivation is to realize superconducting sensor elements with integrating capabilities similar to their CMOS-sensor counterparts. The pixels can operate in several modes. First, we demonstrate that photons can be counted individually, with each detection event adding an identical amount of supercurrent to an integrating element. Second, we demonstrate an active gain control option, in which the signal added per detection event can be dynamically adjusted to account for variable light conditions. Additionally, the pixels can either retain signal indefinitely to record all counts incurred over an integration period, or the pixels can record a fading signal of detection events within a decay time constant. We describe additional semiconductor readout circuitry that will be used in future work to realize scalable, large-format sensor arrays of superconducting single photon detectors compatible with CMOS array readout architectures.
\end{abstract}


\maketitle

Superconducting-nanowire single-photon detectors (SPDs) are gaining maturity with reported system detection efficiencies greater than 98\,\% \cite{Reddy:2020} and timing jitter below 3\,ps \cite{Korzh:2020}. These devices can detect single quanta of radiation, and the same basic device concept can be used from the UV to mid-IR \cite{Verma:2021}, with recent demonstrations showing high efficiency up to 10\,\textmu m wavelength\cite{Verma:2021}. SPDs have very low dark counts and can be fabricated with a simple process into relatively large (400 kilopixel) arrays \cite{oripov:2023}. Large arrays are desirable for many purposes such as imaging and spectroscopy with applications including astronomy \cite{Wollman:2021}, semiconductor circuit metrology \cite{MCMANUS:2000,Stellari:2019}, and biomedical imaging \cite{Ozana:2021,Emrich:2023}. However, readout of large arrays remains the primary impediment to adoption in deployed systems. When an SPD detects a photon, the bias current is diverted from the wire. Room temperature amplifiers and digital electronics are typically used to read out the current pulses from each detector independently. This approach often requires large numbers of coaxial cables inside the cryostat and extensive room-temperature electronics, limiting scaling to larger arrays. Additionally, when detecting photons deeper in the mid-IR, the energy of the photons decreases, necessitating narrower nanowires, which carry less current and provide smaller pulses for detection. Directly amplifying and measuring each diverted current pulse is not an ideal measurement technique for large arrays of SPDs. For the largest arrays demonstrated to date \cite{oripov:2023}, multiplexed readout lines are employed that limit total system count rates, and the means of encoding SPD detection events onto the readout line requires large SPD currents, which makes operation with mid-infrared SPDs difficult.

Several approaches to readout of SPDs \cite{Steinhauer:2021} involve the transduction of pulses from the SPD to supercurrent using superconducting electronic circuitry. Example approaches include using circuitry based on nanowires \cite{Onen:2020} or Josephson junctions (JJs). When using JJs, it is possible to leverage superconducting quantum interference devices (SQUIDs) as sensitive flux-to-voltage transducers \cite{Kirste:2009} or to make use of digital processing with single-flux-quantum circuits \cite{Miyajima:2017,Miki:2018,Miyajima:2018,Yabuno:2020} or adiabatic quantum flux parametron circuits \cite{Takeuchi:2020,Terai:2012,Miyajima:2019,Ortlepp:2011}. Most of these efforts using JJs to readout SPDs have employed separate chips for the sensors and the readout electronics with wire bonds between the specialized die; two recent demonstrations have accomplished monolithic integration of SPDs with JJs \cite{Khan:2022,Miyajima:2023}.

Here we propose and demonstrate an approach to large-scale SPD array readout that makes use of SPDs integrated with JJs and SQUIDs to introduce a technique of photon count integration akin to CMOS cameras. We demonstrate single-photon integrating pixels suitable for this approach wherein the history of photon detection events is locally stored at each pixel, analogous to charge accumulation in CMOS sensors \cite{Fossum:2014}. Circuits performing the desired functions can be achieved through the monolithic integration of SPDs with JJs. In these circuits, an SPD works in conjunction with JJ circuits to transduce photon detection events into current that can be stored indefinitely in a superconducting loop at each pixel. This approach decouples detection events from the readout process. Low-noise readout can be accomplished through a measurement duration that is not limited by the temporal extent of the SPD current pulse. This measurement can be accomplished with MOSFET circuits that transduce the integrated current signal to charge on a capacitor. At that point, the readout proceeds exactly as in a CMOS sensor array. SPD-JJ integration circumvents the small SPD current signal in the mid-IR regime, as explained later. Further integration with MOSFETs enables a low-noise, scalable readout framework that never misses an SPD count, only sends the required information to room temperature, and transduces low-voltage superconductor signals to semiconductor-level voltages to be processed by conventional silicon electronics. This integrated approach leverages the best attributes of superconducting sensors with the convenience and fieldability of semiconductor array readout concepts. In this work we demonstrate the first processing stage of this concept where the SPD signal is transduced to integrated supercurrent by JJ circuits. The further integration with MOSFETs will be the subject of future work.

\begin{figure}
\includegraphics[width=\linewidth]{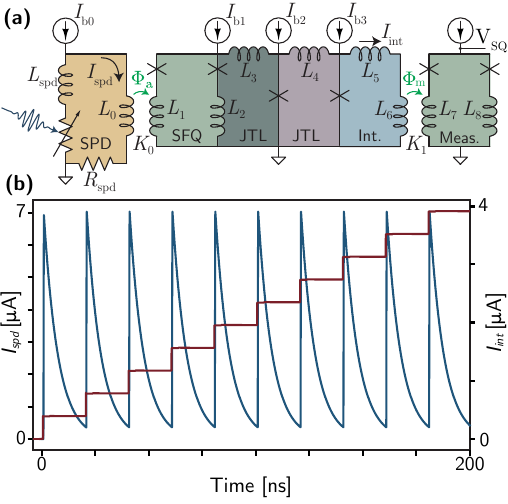}
\caption{\label{fig:1stSchematic}(a) Schematic of superconducting single photon counter. Parameters used in simulation and design are same, except $R_\mathrm{spd}$ which was increased from 12.375\,$\Omega$ to 123.75\,$\Omega$ in simulation, to reduce simulation time. Rest of the parameters were $L_\mathrm{spd}=825$\,nH, $L_{0}=3.1$\,nH, $L_{1}=12.7$\,pH, $L_{2}=1.1$\,pH, $L_{3}=L_{4}=10.34$\,pH, $L_{5}=5.3$\,nH, $L_{6}=0.18$\,nH, $L_{7}=9.36$\,pH, $L_{8}=0.8$\,pH. Mutual inductance couplings were $K_{0}=0.5$ and $K_{1}=0.25$.  Josephson Junction's critical currents $I_{c}$ were 100\,$\mu$ A. On the other hand, bias currents in the simulation were $I_\mathrm{b0}=10 \mu A$, $I_\mathrm{b1}=140 \mu A$, $I_{b2}=I_{b3}=70 \mu A$. (b) Circuit simulation. The blue curve is the current diverted from the SPD ($I_\mathrm{spd}$), which is plotted relative to the left $y$-axis. The red curve is the current being integrated in response to each SPD pulse ($I_\mathrm{spd}$), which is plotted relative to the right $y$-axis.}
\end{figure}

The single-photon integrating pixel concept demonstrated here is shown in Fig.\,\ref{fig:1stSchematic}. A detection event from the SPD is converted into a single-flux quantum (SFQ) through an inductively-coupled DC-SFQ converter \cite{van:1981, Kadin:1999}. The SFQ pulse leaving the DC-SFQ converter then propagates down a short Josephson transmission line and is added to an integration loop at the pixel. The Josephson transmission line comprises two junctions between the DC-SFQ converter and the integration loop, as shown in Fig.\,\ref{fig:1stSchematic}(a). One benefit of this approach is that the input to the DC-SFQ is not current, but rather magnetic flux, which is the product of current and mutual inductance. The mutual inductance between the SPD and the DC-SFQ can be quite large, limited primarily by space, enabling even small SPD pulses to generate SFQ pulses. SPD output currents can be as small as 100\,nA when detecting long-wavelength photons in the mid-IR.  The price is area, with the DC-SFQ receiving the SPD pulse requiring an area of 30\,\textmu m x 30\,\textmu m. Still, a megapixel array would fit on a sensor chip 3\,cm x 3\,cm. Therefore, by integrating JJs with SPDs we can overcome the limitation of small current signals and provide local integration of the signal at each pixel. Timing information is retained at the level of the 10\,kHz frame rate (100\,\textmu s) as opposed to the sub-ns jitter of the detection event. The retained temporal information is more than sufficient for imaging and spectroscopy. By separating photon detection and integrated signal measurement, we can integrate photon counts for as long as desired and separately measure the accumulated signal for as long as necessary to realize noiseless readout.

The circuit in Fig.\,\ref{fig:1stSchematic}(a) involves an SPD, a DC-to-SFQ converter (topologically equivalent to a DC SQUID), a Josephson transmission line, an integration loop that stores the pulses, and a readout component. A readout SQUID is used in this report as the readout component, but in the final scheme a row-column bus architecture identical to CMOS sensor arrays is envisioned, as discussed next. The first step in the circuit operation is the detection of a photon by the SPD. When a photon that is absorbed by the SPD breaks superconductivity and causes a resistive hot spot \cite{natarajan2012superconducting}, the bias current $I_\mathrm{b0}$ will be diverted from the SPD into a transformer coupled to the SPD-to-SFQ SQUID, labelled as SFQ in Fig.\,\ref{fig:1stSchematic}(a). Simulation of the SPD current pulses, $I_\mathrm{spd}$, are shown in  Fig.\,\ref{fig:1stSchematic}(b) by the blue curve. The next step is the transduction of each detected photon into an individual fluxon. A fluxon is a quantum of magnetic flux denoted by and equal to $\Phi_0 \equiv h/2e$ $\approx 2\,\mathrm{mV}\cdot \mathrm{ps}$. Each fluxon generated by the SPD-SFQ transducer circuit propagates through a Josephson transmission line and is stored as current in an integration loop. In the present case, the SPD provides flux input after each photon detection event, producing a discrete amount of supercurrent that is stored in an inductive loop after passing through the Josephson transmission line, which we refer to as the detector integration (DI) loop. In response to a single photon detection event, the integrated current can either be in the form of a single flux quantum or it can be in the form of several flux-quantum pulses. In the former case we have digital operation with each photon detection event producing an identical current signal, which we refer to as SPD-SFQ operation or the digital mode. In this case, the amount of current added to the loop per photon detection is  $\Phi_0/L$, where $L$ is the loop inductance. Current in the integration loop is shown by the red curve in Fig.\,\ref{fig:1stSchematic}(b). In the latter case each photon leads to an analog signal that can be adjusted with a control bias ($I_\mathrm{b1}$). The main advantage of the digital mode of operation is that it is conducive to zero-noise operation and less dependent on circuit bias current, $I_\mathrm{b1}$. There is a one-to-one correspondence between the number of detected photons and stored fluxons. This results in a near-perfectly linear relationship between photon count and integrated signal. The amplitude of added signal accumulated with each photon detection event is identical across an extremely broad range. With this mode of operation, 1024 photon pulses can be detected with exactly linear response across all pulses. On the other hand, the advantage of the analog mode is that the amplitude of the signal can be adapted by adjusting the bias current, $I_\mathrm{b1}$, which enables gain control for adjustable performance depending on the light level. The same hardware infrastructure supports both modes of operation.

In either the analog or digital case, the generated current is stored in the integration loop where the signals from repeated photon-detection events are summed. With the integration loop there are again two modes of operation. In one mode the loop has zero resistance, and the signal is stored with no decay, as shown in Fig.\,\ref{fig:1stSchematic}(b). In this mode of operation, the current stored in the loop immediately preceding a read event is proportional to the total number of photons that have been detected in that integrate-read cycle. In the second mode, the integration loop has a finite resistance, and the integrated signal leaks with a rate given by the loop $\tau = L/r$ time constant \cite{Khan:2022}. The signal stored in the pixel is proportional to the rate of photon detection events in the preceding time interval of order $\tau$. In this mode of operation, no external signal is required to reset the state of the loop. This is a leaky integrator wherein the signal is proportional to the recent rate of photon detection events, so in this mode the sensor is a watt meter. The exact same pixel can be used in either power or energy integration mode with the inclusion of a simple resistive element.

\begin{figure}
\includegraphics[width=\linewidth]{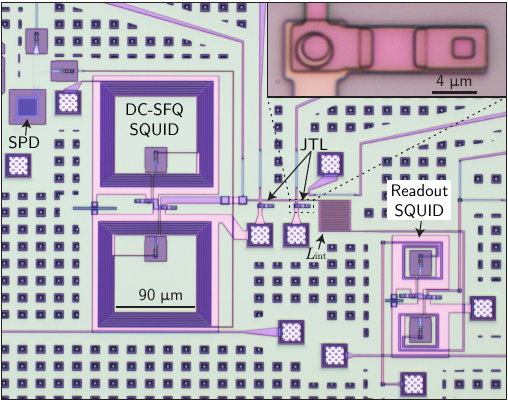}
\caption{\label{fig:uscope} Microscope image of the fabricated device.}
\end{figure}

The fabrication consists of fifteen mask layers. Electron-beam lithography was used for the SPD step, while all other patterning was accomplished with photolithography using a 365\,nm $i$-line stepper. A complete process flow can be found in Ref. \onlinecite{Khan:2022}. In brief, a 40\,nm Nb wiring layer for contact to the SPDs was patterned using a liftoff process. Liftoff was used to avoid a vertical edge and provide a gradual, sloping contact for the thin film used for the SPDs. The SPDs were formed from a 4.1\,nm thick MoSi film \cite{Lita:2021}, which was sputtered after the Nb contact layer. The MoSi was patterned into a detector meander using electron-beam lithography to realize wire widths around 200\,nm. An interlayer dielectric of $\mathrm{SiO_2}$ insulates the SPD layer from a Nb ground plane above it. The JJ trilayer stack (Nb-aSi-Nb) \cite{Olaya:2019} is then deposited and patterned above the ground plane, with another SiO$_2$ insulator in between. PdAu resistors are patterned and deposited with liftoff to form the JJ shunt resistors. An additional low-resistance Au layer was used to make resistors with small values and therefore long attainable leak time constants in the integrating loops used in power-meter mode, described next. An additional top insulator sealed the structures. All layers were connected with Nb vias through the insulators. Microscope images of the fabricated device are shown in Fig.\,\ref{fig:uscope}.

\begin{figure*}
\includegraphics[width=\linewidth]{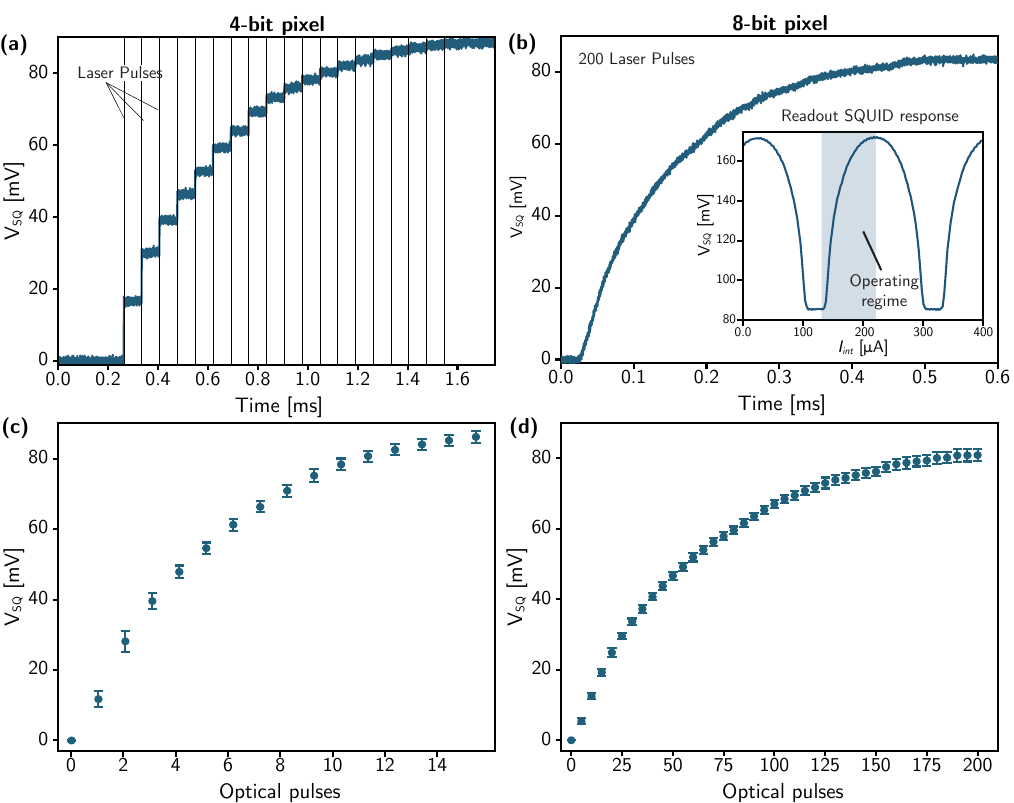}
\caption{\label{fig:response}(a) Response of 4\,bit pixel to a train of input optical pulses. (b) Response of 8\,bit pixel. Inset shows the readout SQUID response curve. Gray region shows the operating regime. (c) Statistical analysis of the pixel measured in (a). (d) Statistical analysis of the pixel measured in (b).}
\end{figure*}

Measurements were performed at 2.3\,K in a closed-cycle Gifford-McMahon cryostat. The chip was flood illuminated by a fiber-coupled, 780\,nm pulsed laser source. The laser pulse width was approximately 480\,ps, while the SPD recovery time was around 37.5\,ns. Therefore, multiple detection events per pulse were unlikely. The maximum voltage from the readout SQUID ($V_\mathrm{SQ}$) was on the order of 10\,\textmu V, and a room-temperature amplifier with 60\,dB voltage gain was used for measurements. A diagram of the measurement apparatus is shown in the supplementary information of Ref. \onlinecite{Khan:2022}.

Figures \ref{fig:response}(a) and (b) show the voltage across the SQUID as a function of time while optical pulses are directed at the SPD at a fixed rate. In Fig.\,\ref{fig:response}(a) the integration loop inductance is 330\,pH, chosen to store 16 pulses (4\,bits), while that of Fig.\,\ref{fig:response}(b) is 5.3\,nH, chosen to store 256 pulses (8\,bits). Discrete steps are evident with each laser pulse in Fig.\,\ref{fig:response}(a) as a fluxon enters the integration loop. These are identical fluxon pulses, but here the measured response is nonlinear because it is convoluted with the response of the readout SQUID. The readout SQUID response is shown in the inset of Fig.\ref{fig:response}(b). The same discrete steps are present in the response of Fig.\,3(b) but are not discernible as they are smaller than the noise, which in this case is due to line noise coupled from our cryostat compressor to our measurement electronics. The full readout scheme, described next, would eliminate both the non-linearity and the noise. Figures \ref{fig:response}(c) and (d) show statistical analysis of the small- and large-storage-capacity loops, respectively. The data points are the SQUID voltage averaged over 1000 independently measured traces. Each trace was taken after the number of photonic pulses indicated on the $x$-axis. The error bars give the standard deviation calculated from the 1000 traces. After each trace was generated and recorded, the current in the integration loop was erased by driving current through a PdAu resistor, fabricated in closed vicinity to the inductor in the DI loop, $L_\mathrm{int}$. To reset the state of the DI loop, a 10\,mA current was applied to the resistor, which heated the inductor, broke superconductivity, and purged the integrated current in the loop.

\begin{figure}
\includegraphics[width=\linewidth]{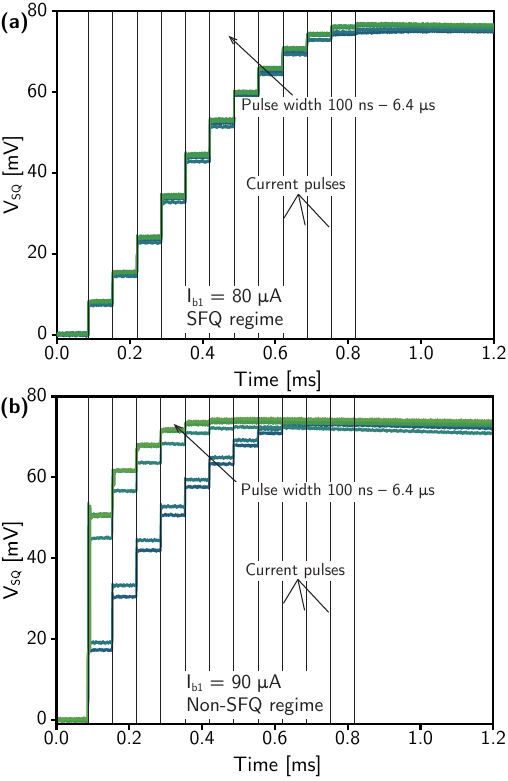}
\caption{\label{fig:SFQ} Photon counter response for different input pulse width (a) 80 µA SFQ SQUID bias (b) 90 µA SFQ SQUID bias. The tiny decrease of current with time is due to a DC block filter.}
\end{figure}

To make sure the photon counters are working in SPD-SFQ mode in Fig.\,\ref{fig:response}, we drove the SPD with current pulses that exceeded the switching current instead of relying on optical pulses. In this way we could change the width of the input pulses. An SPD-SFQ converter should output only one SFQ pulse for each input pulse, regardless of the duration of the input pulse. Figure \ref{fig:SFQ}(a) shows the device response for the current pulse width varying from 100\,ns to 6.4\,\textmu s, in steps of a factor of two (geometrically spaced). No significant change in the output of the device is observed. The inductor in the integration loop can hold only around 16 SFQ pulses before it saturates. Because it does not saturate at a lower number of pulses when driven with a larger pulse width, we conclude the device is operating in the SPD-SFQ regime. In Fig.\,\ref{fig:SFQ}(b) we increased the bias to the SFQ SQUID, from 80\,\textmu A to 90\,\textmu A, thus moving the circuit outside the SPD-SFQ regime. In this case the integration loop saturates with fewer pulses as we increase the pulse width. Furthermore, the readout SQUID voltage step should be less than $\approx 10$\,mV for a single SFQ pulse, considering the inductance of the integration loop, room temperature amplification, and operating point on the readout SQUID response curve. Hence, at around 80\,\textmu A SFQ-SQUID bias, the device is working in the desired SPD-SFQ mode, while by 90\,\textmu A SFQ-SQUID bias it is no longer producing exactly one fluxon per photon detection event.

\begin{figure}
\includegraphics[width=\linewidth]{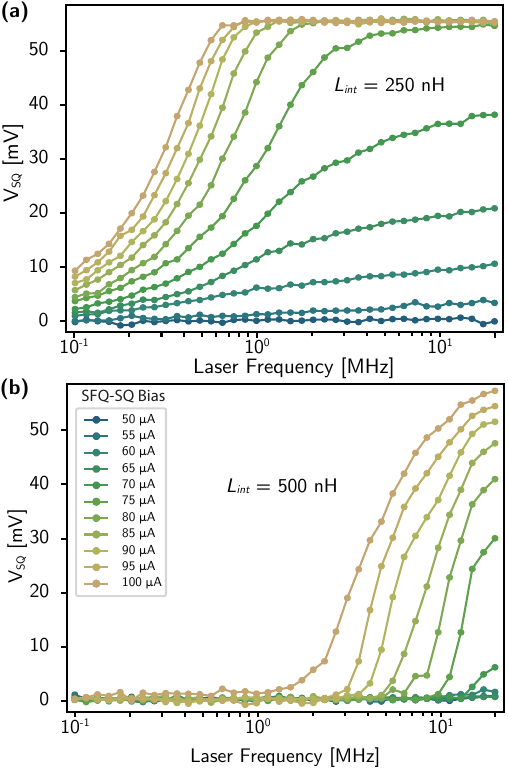}
\caption{\label{fig:powermode} Steady-state SQUID voltage as a function of pulse train frequency at different values of SFQ-SQUID bias, $I_\mathrm{b1}$ (a) DI loop inductance is 250\,nH. (b) DI loop inductance is 500\,nH.}
\end{figure}

In addition to the digital SPD-SFQ operation just described, we have conducted measurements of similar circuits operated in analog, power-meter mode. For this demonstration, a slightly different circuit was used wherein the initial transduction SQUID (referred to as SPD-SFQ up to this point) had more symmetric inductances, compared to the counter ($L_1$ = 9.2\,pH  and $L_2$ = 5.4\,pH in Fig.\,\ref{fig:1stSchematic}(a)). This symmetric design is employed to implement the analog as opposed to digital transduction operation. The modified circuit also included a resistor in the integration loop, providing a leak rate. Thus, the circuit used in this part of the study is an analog power meter as opposed to the digital energy meter demonstrated by the data in Fig.\,\ref{fig:response}. In this mode, the initial transduction SQUID produces a stream of fluxons with each detection event. The number of fluxons generated with each detection event is determined by this bias current $I_\mathrm{b1}$, which provides a control knob to adjust the response of the pixel based on the light level. The decay time here is around 6.25\,\textmu s, which is determined by the $L/R$ decay time of the integration loop. For sufficiently long pulse trains at a given frequency, the device reaches a steady state that can be tuned with $I_\mathrm{b1}$. Figure \ref{fig:powermode} shows the voltage on the readout SQUID as a function of the frequency of input photonic pulses. The different traces correspond to different values of $I_\mathrm{b1}$, and this dynamically variable control parameter can be used to adjust the response to keep the pixel in a useful dynamic range. For a given value of $I_\mathrm{b1}$ and a given input rate of photons, the pixel will reach a different steady state value, which can then be used to determine the incident light flux through a calibration procedure. Figure \ref{fig:powermode} shows the measured SQUID voltage as a function of the incident pulse rate for values of the bias current $I_\mathrm{b1}$ from 50\,\textmu A to 100\,\textmu A, demonstrating tuning across several orders of magnitude in incident photon flux. These are the transfer functions that can be used to determine the rate of incident photons. Fig.\ref{fig:powermode}(a) shows data for the device with 250\,nH inductance in the DI loop, while Fig.\ref{fig:powermode}(b) has an inductance of 500\,nH. We see that with this range of input signal rates and integration-loop leak rate, the dynamic range of the smaller capacity loop is better matched to the signal.

\begin{figure}
\includegraphics[width=\linewidth]{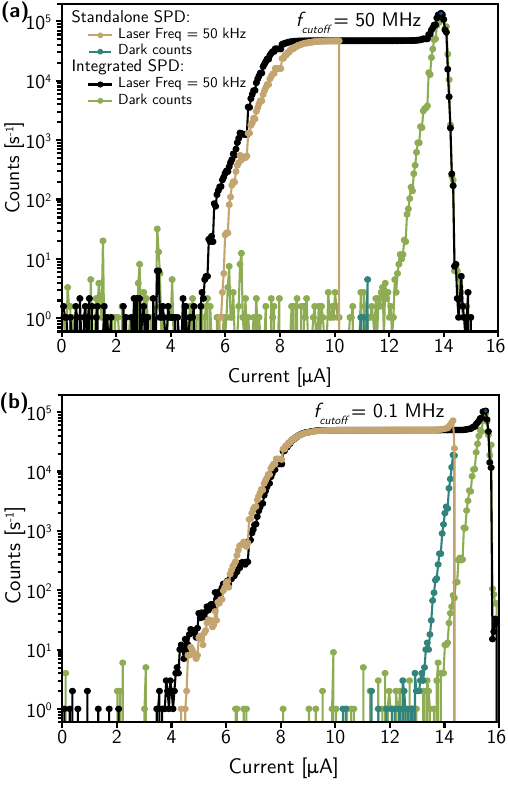}
\caption{\label{fig:enhance} Comparison of readout performance from a standalone and integrated SPD (a) using room temperature amplifier of 50 MHZ lower cutoff frequency (b) using room temperature amplifier of 0.1 MHZ lower cutoff frequency.}
\end{figure}

During our experimentation we discovered that reading SPD pulses through Josephson electronics can reduce the effect of amplifier noise compared to stand-alone SPD readout. In the stand-alone SPD readout scheme, the SPD is connected to an amplifier through a bias tee. If there are high-frequency reflections from the amplifier, they will pass though the bias tee and affect the SPD. This amplifier noise will result in lowering the SPD operating range and limit the device operation to a smaller plateau. On the other hand, in the integrated SPD scheme reported here, the SPD is not connected to a bias tee. It is well isolated from the amplifier though two superconductor SQUID transformers, leading to a more stable SPD bias current.  Figure \ref{fig:enhance} shows the count rate versus SPD bias current for both a standalone and an integrated SPD that were fabricated on the same wafer, with the same geometry, located in close proximity on the chip. For the integrated SPD we used the power-meter device with a leak in the integration loop. For sufficiently low laser frequency, the decay time in the DI loop is smaller than the separation of laser pulses. Therefore, individual voltage pulses were discernable and could be counted using a commercial pulse counter. We used a laser frequency of 50\,kHz for both standalone and integrated SPD. In Fig.\,\ref{fig:enhance}(a), the room temperature amplifier used has a lower cutoff frequency of 50\,MHz, and therefore frequencies below 50\,MHz are expected to reflect from the amplifier. This lowers the operating range of the standalone SPD. On the other hand, the integrated SPD is not affected by the reflection, resulting in a factor of three improvement in the width of the plateau region. Figure \ref{fig:enhance}(b) shows the same experiment but with an amplifier of 0.1\,MHz lower cutoff frequency. Here the plateau region of the standalone SPD comes closer to matching that of the integrated device, yet the integrated SPD still has larger plateau region by roughly 15\%.

We have proposed and demonstrated a single-photon-counting pixel that functionally resembles a CMOS sensor pixel. The concept is made possible by the monolithic integration of SPDs with JJs. We have shown here that such pixels can store the signals from several hundred photon-detection events as supercurrent for later readout. This transduction of photons to stored supercurrent relies on the interface between single photon detectors with DC-SFQ converters. We have shown that the same basic circuit concept can be used to form a leaky integrator pixel that retains information about the average photon flux incident within the a time period set by the leak rate of the integration loop.

\begin{figure}
\includegraphics[width=\linewidth]{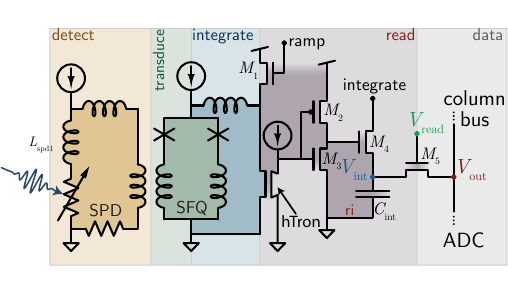}
\caption{\label{fig:pixel} Circuit concept combining SPD, SQUID transduction, integration loop, and CMOS readout.}
\end{figure}

In all modes of operation, the signal of interest is supercurrent in an inductor. To achieve  scalability matching that of CMOS sensor arrays, it is necessary to extend the work here to include a readout architecture that will allow the interrogation of these supercurrents stored in arrays of millions of pixels. While multiple approaches to this technical challenge may be possible, we contend that further integration of SPDs and JJs with CMOS readout electronics offers a uniquely scalable approach to the problem. The integrated current in each pixel can be read out with transistors in an architecture very similar to that of CMOS sensor arrays. To interface superconducting electronics to semiconducting electronics, superconducting signals must be stepped up in voltage. Recent work on superconducting thin film amplifiers has significantly improved semiconductor-superconductor interfaces \cite{Zhao:2017,McCaughan:2019}. In the superconducting state these amplifiers have zero resistance and can carry appreciable currents. When switched to the normal state by a current pulse, they transition to a high resistance state within less than a nanosecond, producing the voltage required to switch a MOSFET. These superconducting amplifiers are referred to as hTrons, and the circuit diagram of Fig.\,\ref{fig:pixel} shows an hTron interfacing the integration portion of the pixel to the CMOS readout circuitry. After an integration period, the detector integration loop contains a current proportional to the number of photons that have been detected. At read time, MOSFET $M1$ provides a ramp that adds to the current through the hTron gate. When $M1$ begins its ramp, $M4$  begins delivering a fixed current to the integration capacitor, $C_\mathrm{int}$. When the sum of the integrated current signal and the applied measurement current from $M1$ reach the hTron gate threshold, a voltage will occur across the hTron channel, switching the gates of $M2$ and $M3$, which form an inverter. When this inverter switches, it cuts the voltage to $M4$, terminating the flow of current to $C_\mathrm{int}$. After this operation, the charge on $C_\mathrm{int}$ is inversely proportional to the current that was present in the integration loop. This charge now serves as a proxy for the number of photons that were detected during integration, with the number of electrons having a direct correspondence to the number of detected photons. With the desired information represented as charge on a capacitor, the remainder of the readout follows a CMOS sensor array exactly. To read the charge on $C_\mathrm{int}$, $M5$ is opened, and the charge is coupled to the column read bus. When the hTron reaches its threshold, the current in the detector integration loop is erased, so the measurement of the state of the loop is destructive, and integration begins again with an empty loop. 

Here we have shown the utility of integrating SPDs with JJs to increase the functionality of each SPD pixel. By further integrating these superconducting components with semiconductor electronics, significant further capabilities in array readout will be enabled. The full semiconductor-superconductor integration will be the subject of future work.

\section*{\label{sec:Ack} Acknowledgements}
This work was funded by the DARPA Invisible Headlights program.

\section*{\label{sec:References} References}
\providecommand{\noopsort}[1]{}\providecommand{\singleletter}[1]{#1}%
%

\end{document}